\documentclass{ws-ijqi}
\renewcommand{\trimmarks}{\relax}

\renewcommand{\draftnote}{\relax}

\def\currenttime{\hour=\time \divide\hour by 60 \number\hour:%
  \multiply\hour by 60 \minute=\time \global\advance\minute by -\hour%
  \ifnum\minute<10 0\number\minute\else\number\minute\fi}

\usepackage{overcite}\let\refcite\citen\relax

\newcounter{fact}
\newenvironment{fact}%
{\refstepcounter{fact}\medskip\begin{center}\begin{minipage}{0.90\textwidth}%
\textbf{Fact~\thefact.}\ \ignorespaces}%
{\end{minipage}\end{center}\medskip}

\newcommand{\ds}{\displaystyle}
\newcommand{\cD}{\mathcal{D}}
\newcommand{\cL}{\mathcal{L}}
\newcommand{\cS}{\mathcal{S}}

\newcommand{\tr}{\mathrm{tr}}
\newcommand{\rhoEst}{\hat{\rho}_{\mathrm{ME}}}
\newcommand{\pkEst}{{\left(\hat{p}_k\right)}_{\mathrm{ME}}}
\newcommand{\Aphy}{{\left(\mbox{\small$\ds\sum$}\nu^2\right)}_{\mathrm{phy}}}

\makeatletter
\DeclareRobustCommand\dyadic[1]{%
\@ontopof{#1}{\leftrightarrow}{1.15}\mathord{\box2}}
\DeclareRobustCommand\roarrow[1]{%
\@ontopof{#1}{\rightarrow}{1.15}\mathord{\box2}}
\renewcommand{\vec}[1]{\roarrow{#1}}

\chardef\f@ur=4\relax
\long\def\true@sw#1#2{#1}%
\long\def\false@sw#1#2{#2}%
\def\@boolean#1#2{%
  \long\def#1{%
    #2% \if<something>
      \expandafter\true@sw
    \else
      \expandafter\false@sw
    \fi
  }%
}%
\def\@boole@def#1#{\@boolean{#1}}%
\@boole@def\@ifdim#1{\ifdim#1}%

\def\@ontopof#1#2#3{%
 {%
  \mathchoice
    {\@@ontopof{#1}{#2}{#3}\displaystyle     \scriptstyle      }%
    {\@@ontopof{#1}{#2}{#3}\textstyle        \scriptstyle      }%
    {\@@ontopof{#1}{#2}{#3}\scriptstyle      \scriptscriptstyle}%
    {\@@ontopof{#1}{#2}{#3}\scriptscriptstyle\scriptscriptstyle}%
 }%
}%
\def\@@ontopof#1#2#3#4#5{%
  \setbox\z@\hbox{$#4#1$}%
  \setbox\f@ur\hbox{$#5#2$}%
  \setbox\tw@\null\ht\tw@\ht\z@ \dp\tw@\dp\z@
  \@ifdim{\wd\z@>\wd\f@ur}{%
    \setbox\f@ur\hb@xt@\wd\z@{\hss\box\f@ur\hss}%
    \mathord{\rlap{\raise#3\ht\z@\box\f@ur}\box\z@}%
  }{%
    \setbox\f@ur\hb@xt@.9\wd\f@ur{\hss\box\f@ur\hss}%
    \setbox\z@\hb@xt@\wd\f@ur{\hss$#4\relax#1$\hss}%
    \mathord{\rlap{\copy\z@}\raise#3\ht\z@\box\f@ur}%
  }%
}%

\makeatother
\newcommand{\JournalTitle}[1]{\textit{#1}\ }

\newcommand{\PRA}{\JournalTitle{Phys.\ Rev.\ A}}

\newcommand{\JMO}{\JournalTitle{J.~Mod.\ Opt.}}
\newcommand{\JMaPh}{\JournalTitle{J.~Math.\ Phys.}}

\newcommand{\Eprint}[3][quant-ph]{%
\mbox{e-print} arXiv:#2\linebreak[0][#1] (#3)}

\begin{document}

\markboth{H.K. Ng and B.-G. Englert}%
{A simple minimax estimator for quantum states}

\title{\uppercase{A simple minimax estimator for quantum states}}

\author{\uppercase{Hui Khoon Ng}}
\address{Centre for Quantum Technologies, %
National University of Singapore\\
3 Science Drive 2, Singapore 117543, Singapore\\
and Applied Physics Lab, DSO National Laboratories\\ 
20 Science Park Drive, Singapore 118230\\
cqtnhk@nus.edu.sg}

\author{\uppercase{Berthold-Georg Englert}}
\address{Centre for Quantum Technologies, %
National University of Singapore\\
3 Science Drive 2, Singapore 117543, Singapore\\
and Department of Physics, National University of Singapore\\
2 Science Drive 3, Singapore 117542, Singapore\\
cqtebg@nus.edu.sg}

\maketitle  

\begin{history}
Received 16 March 2012
\end{history}

\begin{abstract}
Quantum tomography requires repeated measurements of many copies of the
physical system, all prepared by a source in the unknown state. 
In the limit of very many copies measured, the often-used maximum-likelihood
(ML) method for converting the gathered data into an estimate of the state
works very well. 
For smaller data sets, however, it often suffers from problems of rank
deficiency in the estimated state. 
For many systems of relevance for quantum information processing, the
preparation of a very large number of copies of the same quantum state is
still a technological challenge, which motivates us to look for estimation
strategies that perform well even when there is not much data. 
After reviewing the concept of minimax state estimation, we use
minimax ideas to construct a simple estimator for quantum states. 
We demonstrate that, for the case of tomography of a single qubit, our
estimator significantly outperforms the ML estimator for small number of
copies of the state measured. 
Our estimator is always full-rank, and furthermore, has a natural dependence
on the number of copies measured, which is missing in the ML estimator. 
\end{abstract}

\keywords{Quantum tomography; state estimation; minimax; maximum likelihood; Bayesian}

\section{Introduction}
Tomography is the art of estimating the state of a system put out by a given
source. 
For example, one might be interested in characterizing the polarization of a
photon from a laser source; or two parties in a communication protocol want to
know the state they jointly receive from a common source; or an experimentalist
might want to verify that a source built in his lab to provide some target
state is indeed meeting its specifications. 
The scope of tomography can be broadened to include parameter estimation,
that is, to estimate a certain quantity of interest for some operational task
(the fidelity between output state and target state, for example, or the
expectation value of some fixed observable on the state). 
In this work, however, we will only deal with the most often discussed case of
estimating the full state. 

Tomography involves two steps: (i) the measurement of identical copies of the
state; (ii) the conversion of the data collected from the measurement into an
estimator for the state. 
In the simplest case, the measurement step (i) involves the same measurement
on every copy of the state. 
More generally, it can be adaptive, that is, the measurement to be made on the
$k$th copy can depend on information gathered from measuring the previous
$k-1$ copies. 
In the estimation step (ii), the simplest method gives a 
\emph{point estimator}, which is a single state that represents our best guess
of the identity of the true state. 
More generally, one can give a set of states compatible with the observed data
that includes the true state with high probability. 
Such \emph{region estimators} are known in classical estimation theory, and
have appeared recently in the quantum
arena~\cite{Christandl11,BlumeKohout12}. 
In our work, we will discuss the simplest case of repeated (non-adaptive)
measurements, particularly measurements with the property of being symmetric
and informationally complete, and focus on the issue of providing a point
estimator. 

The most popular procedure leading to a point estimator goes under the
collective name of \emph{maximum-likelihood} (ML) methods, first applied to
quantum tomography by Hradil~\cite{Hradil97}. 
ML methods prescribe as the point estimator the state with the largest
likelihood of giving rise to the observed data, and there are numerous
variations to this theme depending on the scenario in question (see
Ref.~\refcite{MaxLikRev} for a good review). 
ML methods are particularly attractive because they do not require the choice
of a prior distribution, a problem that plagues alternative methods based on
Bayesian ideas (see, for example, Refs.~\refcite{Cousins95} and
\refcite{Kass96}). 
In the limit of a very large number of copies of the state measured, ML
methods work very well---the likelihood function becomes so sharply peaked
around the true state that one requires little sophistication to make a good
guess. 

However, for small sample sizes, ML methods are perhaps less well-motivated,
and there is reason, as we will see in Section \ref{sec:meanEst}, to look for
other methods in the estimation step. 
Besides, one should hardly expect that ML methods are the best choice for all
scenarios, since its optimality is based on a particular figure-of-merit,%
\footnote{Here is a comment that will likely make sense to the reader only
  upon reading the remainder of the paper: 
  The ML estimator can be shown to be optimal in terms of minimizing the
  average risk, where the averaging uses the prior distribution
  $\mathrm{d}\mu=\mathrm{d}p_1\mathrm{d}p_2\ldots \mathrm{d}p_K$ for a state
  characterized by probabilities $\{p_k\}_{k=1}^K$. 
  The estimation error is quantified by a cost function that assigns a value
  $0$ only when the estimator and the true state are identical, and $1$
  otherwise (see, for example, Ref.~\refcite{Jaynes}). 
  Hence, even though the ML estimator requires no choice of prior distribution
  in its construction, in judging its efficacy, one still requires a choice of
  prior distribution to quantify its average performance over all true
  states. 
  This point will be reiterated later in the text.} 
and whether this is a suitable figure-of-merit will undoubtedly depend on the
task at hand. 
This motivates us to look beyond ML methods for alternative strategies
appropriate for different tasks. 

A class of estimation procedures that requires no arbitrary or subjective
choice of prior distribution is the class of \emph{minimax} methods. 
In a minimax procedure, one looks for an estimator, usually from within a
specified class of estimators, that gives the smallest worst-case (over all
true states) estimation error. 
This gives an optimality condition that holds regardless of the probability of
occurrence of each true state.
While such a ``worst-case scenario'' approach may be overly cautious for 
some purposes, it can be suitable and, in fact, necessary for tasks like
cryptography where one would prefer to acknowledge ignorance rather than make
a wrong guess. 

Minimax procedures are, unfortunately, notoriously difficult to implement,
even for classical problems. 
This is hardly surprising since they involve a double optimization---first a
maximization of the estimation error over all possible true states, followed
by a minimization of this maximum over the class of estimators under
consideration.  

However, if we employ the commonly used \emph{mean squared error} to quantify
the estimation error, a minimax estimator with particularly nice features is
known for the problem of a $K$-sided classical die. 
While the minimax estimator for the quantum analog of this problem is not
known, we demonstrate here a general procedure to obtain a point estimator for
the quantum problem that retains most of the desirable features of the
classical minimax solution. 
This estimator is not minimax in the set of all estimators, as is the case for
its classical analog,
but is minimax within a smaller class of estimators with mathematical
structure motivated by the solution for the classical die problem. 
This quantum generalization of the minimax point estimator, despite being
rather ad-hoc in its construction, performs remarkably well in comparison to
ML estimators for the qubit case investigated in detail. 
Furthermore, the estimator is easy to use as it requires no complicated
numerical optimization. 
It can find utility as a good first guess for tomographic experiments,
particularly if one only has access to a small number of copies of the state. 
Applying a similar procedure to adapt other known estimators for the classical
problem to the quantum case might be equally fruitful. 

Our goal here is partly to review the use of minimaxity as a means of choosing
an estimation procedure. 
This is, of course, well known in the classical estimation theory community. 
In the quantum context, however, while minimax ideas have appeared in the
quantum state estimation literature (see, for example,
Refs.~\refcite{MinimaxQ1,MinimaxQ2,MinimaxQ3}), they remain little explored. 
Here, we organize the ideas into a consistent programme (Section
\ref{sec:Minimax}), and contribute by presenting a simple estimator for
quantum states motivated by minimax considerations 
(Section~\ref{sec:Quantum}). 
The geometrical properties of symmetric quantum measurements are discussed in
an appendix, and two more appendices contain mathematical details.

\section{Minimax estimation}\label{sec:Minimax}
We first review two types of estimators---maximum-likelihood and mean
estimators---before leading up to the idea of minimaxity.  
We also review the well-studied problem of the classical die, which serves two
purposes: first, to define the notation and provide a concrete example for the
application of the different estimation procedures; second, to provide
guidance in the quantum problem studied in the next section. 
The reader is to note, however, that the ideas of state estimation discussed
in this section are equally applicable to the quantum problem. 
In moving to the quantum arena, there are significant differences in the setup
of the problem that complicate the application of the state estimation
procedures, but the ideas behind each procedure remain unchanged. 
This is a reminder that the classical state estimation literature has much to
teach us, even in the quantum context.

\subsection{The classical $K$-sided die}
Consider a $K$-sided die with faces labeled $k=1,2,\ldots,K$. 
The probability that face $k$ turns up when the die is tossed is denoted by
$p_k$. 
Tosses of the die are described by the probability distribution
$\{p_k\}_{k=1}^K$, with $p_k\geq 0$ and $\sum_k p_k=1$. 
Suppose we are given a die for which the probabilities are unknown, and we are
allowed $N$ tosses of that die to attempt an estimate of the $p_k$ values.

Let us discuss the tomography of a $K$-sided die using language suitable for
quantum state tomography. 
We write the state of a die with probability distribution $\{p_k\}_{k=1}^K$ as
$\rho=\sum_{k=1}^K|k\rangle p_k \langle k|$, where ket $|k\rangle$ represents 
face $k$ turning up in a toss of the die. 
We can think of $\{|k\rangle\}$ as a basis for the state space with an inner
product defined such that $\langle k|l\rangle=\delta_{kl}$. 
A single toss of the die is then, in this language, a measurement in the basis
$\{|k\rangle\}_{k=1}^K$.  

We can describe this measurement formally as a 
\emph{probability operator measurement} (POM) 
with outcomes $\Pi_1,\Pi_2,\ldots,\Pi_K$. 
To define a POM, the operators $\Pi_k$ must be non-negative and normalized to
unit sum,
\begin{equation}
\Pi_k\geq 0 \quad\textrm{for all } k,\qquad\textrm{with} \quad\sum_k\Pi_k=1.
\end{equation}
The measurement can be thought of as comprising $K$ detectors, each
corresponding to one of the POM outcomes $\Pi_k$. 
The probability that the $k$th detector clicks, if we have the input state
$\rho$, is given by Born's rule 
\begin{equation}\label{eq:BornsRule}
p_k=\tr{\left\{\Pi_k\rho\right\}}.
\end{equation}
For the case of the $K$-sided die, the POM corresponding to a single toss of
the die can be described using the POM $\{\Pi_k\equiv |k\rangle\langle k|\}$,
and $p_k$ is simply the probability that the face $k$ turns up in a single
toss of the die. 

To improve the efficiency of the tomographic measurement, one often chooses a
POM that is \emph{symmetric} (S). 
The properties of S-POMs are the subject of \ref{app:Geometry}.
Here, we are content with considering S-POMs that have rank-$1$ outcomes, in
which case
\begin{equation}\label{eq:S}
\tr\{\Pi_k\Pi_l\}
=\frac{d^2}{K^2}\biggl[\delta_{kl}+\frac{K-d}{(K-1)d}(1-\delta_{kl})\biggr]
\quad\text{for all }k, l,
\end{equation}
where the dependence on $k$ and $l$ is in the Kronecker deltas only.
In particular, for ${K=d}$, this covers the case of the classical die, for
which this S-POM is informationally complete (IC).
One then speaks of a SIC-POM.%
\footnote{Here, we take the liberty to include the classical case under the
  general name of SIC-POM. 
  In the classical case, the SIC-POM is simply a projective measurement---a
  von Neumann measurement. 
  A projective measurement is IC for the classical case, though not for the
  quantum case.}  

Given a SIC-POM, every state of the system can be written as
\begin{equation}\label{eq:invertBorn}
\rho=\sum_{k=1}^K p_k\Lambda_k,
\end{equation}
where $p_k$ is computed via the Born's rule of \eqref{eq:BornsRule} for
the SIC-POM, and the $\Lambda_k$s are hermitian, unit-trace operators
with ${\tr\{\Pi_k\Lambda_l\}=\delta_{kl}}$ as their defining property.
Specifically, we have  
\begin{equation}\label{eq:Lambda}
\Lambda_k\equiv \frac{(K-1)K}{(d-1)d}\biggl[\Pi_k-\frac{K-d}{(K-1)K}\biggr] 
\end{equation}
for the $\Pi_k$s of \eqref{eq:S}.
For the $K$-sided die, $\Lambda_k=\Pi_k$. 
Equation \eqref{eq:invertBorn} can be thought of as inverting Born's rule,
that is, we can write down the state $\rho$ that will give rise to the
probabilities $p_k$ via Born's rule for the SIC-POM. 
The set of probabilities $\{p_k\}$ thus provides a complete description of the
state $\rho$---this is the sense in which the SIC-POM is informationally
complete. 
We will often use the notation $\rho\sim\{p_k\}$ to denote this relation. 
We will also occasionally use simply $p$ to denote the list
$\{p_1,p_2,\ldots,p_K\}$. 
Any set of probabilities $\{p_k\}$ is always understood to satisfy $p_k\geq 0$
for all $k$ and $\sum_kp_k=1$, and we sometimes refer to the set as a
probability distribution. 

The goal of a tomographic problem, classical or quantum, is to provide a
reasonable estimator for the true state $\rho$, given the data from performing
a chosen IC-POM on every one of $N$ identical copies of the input state. 
To be concrete, in the subsequent analysis, we represent the data from the $N$
measurements as a sequence of clicks in the $K$ possible detectors: 
$D_N\equiv\{c_1,c_2,\ldots,c_N\}$, where $c_l\in\{1,2,\ldots,K\}$ is the
detector that clicked in the $l$th measurement. 
We can summarize the data by collecting together the number of clicks for each
detector: $\{n_1,n_2,\ldots,n_K\}$, where $n_k$ is the number of times the
$k$th detector clicked in the $N$ measurements. 
We will use the notation $D_N\sim\{n_1,n_2,\ldots,n_K\}$ to refer to the
summary of a particular sequence of measurement outcomes. 
Note that the data must satisfy $\sum_{k=1}^K n_k=N$. 

A point estimator $\hat{\rho}$ is a map from the set 
$\cD\equiv \{\cD_1,\cD_2,\ldots,\cD_N,\ldots\}$ of all possible data to the
set $\cS$ of all possible (physical) states. 
Here, $\cD_N$ denotes all possible measurement outcomes on $N$ copies of the
state. 
For the classical die problem, for example, $\cD_N$ consists of all possible
sequences of faces revealed in $N$ tosses. 
The set $\cS$ consists of all states $\rho=\sum_kp_k\Lambda_k$ where $\{p_k\}$
is a probability distribution. 
We denote the point estimator for data $D_N$ as $\hat{\rho}(D_N)$, and denote
the set of all point estimators, that is, all maps $\hat\rho:\cD\rightarrow\cS$,
by $\hat\cS$.

\subsection{The maximum-likelihood estimator}
A very popular approach to a point estimator is the maximum-likelihood
method. 
The ML method prescribes as the point estimator the state at which the
likelihood function for the observed data attains its maximum. 
The likelihood function---the probability that the state $\rho\sim\{p_k\}$
gives rise to the data $D_N$---is 
\begin{equation}\label{eq:llhDie}
\cL(D_N|\rho)=\prod_{k=1}^K p_k^{n_k}.
\end{equation}
To find the ML estimator for the classical die problem, we solve the following
constrained maximization problem: 
\begin{align}
&\max_{\rho\sim\{p_k\}} \cL(D_N|\rho),\nonumber\\
\textrm{subject to }\quad&\sum_k p_k=1 
\quad\textrm{with}\quad p_k\geq 0~\textrm{for all } k.
\end{align}
This gives the ML estimator 
$\hat{\rho}_{\mathrm{ML}}\equiv \sum_k\left(\hat{p}_k\right)_{\mathrm{ML}}\Pi_k$ with
\begin{equation}\label{eq:MLdie}
\left(\hat{p}_k\right)_{\mathrm{ML}}=\frac{n_k}{N}\equiv\nu_k.
\end{equation}

\begin{figure}[t]
\begin{center}
\includegraphics[width=0.6\textwidth]{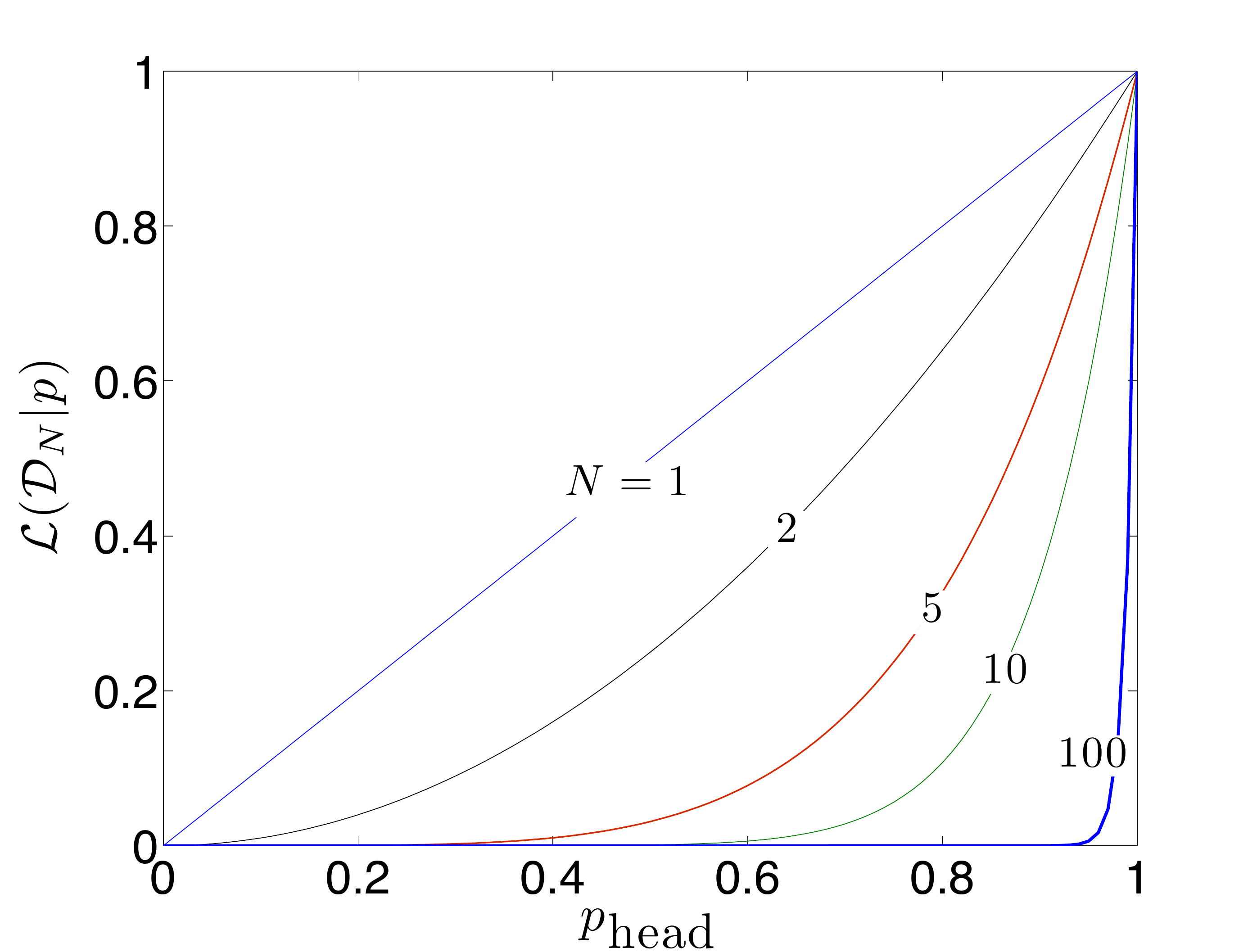}
\caption{\label{fig:llhN}%
Likelihood function for a 2-sided die (coin) with data $D_N\sim\{N,0\}$, 
for $N=1,2,5,10$ and $100$.}
\end{center}
\end{figure}

For large $N$, the ML estimator for the classical die is intuitive: 
From a frequentist's perspective, the long-run ($N\rightarrow\infty$) relative
frequencies $\nu_k$ should approach the probabilities $p_k$. 
What about small $N$? 
Suppose a coin (a ``2-sided die'') is tossed just once, and gives ``heads''. 
Hardly anyone will put his money on the estimator $\hat{p}_k=n_k/N$, which
means setting $\hat{p}_{\mathrm{head}}=1$, and $\hat{p}_{\mathrm{tail}}=0$. 
This lack of confidence in the estimator is well justified if one considers
the fact that, for $D_1\sim\{1,0\}$, the likelihood function is not very
sharply peaked at $p_{\mathrm{head}}=1$, and $p_{\mathrm{head}}$ values near
$1$ have similar likelihood. 
Suppose we make more tosses, and always get heads. 
Then, we gain confidence in the estimator $\hat{p}_{\mathrm{heads}}=1$ and 
$\hat{p}_{\mathrm{tails}}=0$, as is reflected by the likelihood function
getting more and more sharply peaked at $p_{\mathrm{head}}=1$; 
see Fig.~\ref{fig:llhN}. 

Observe that the ML estimator in \eqref{eq:MLdie} depends only on the
relative frequencies $\nu_k$, and not on $N$, the total number of tosses
made. 
For the above example where $D_N\sim\{N,0\}$, the ML estimator is always
$\hat{p}_{\mathrm{heads}}=1$ and $\hat{p}_{\mathrm{tails}}=0$ for all $N$. 
Only the confidence (loosely quantified by the width of the likelihood
function) in the estimator changes with $N$. 
In many situations, only the point estimator, and not the confidence interval
associated with it, is carried forward into subsequent analysis. 
However, a statement that $\hat{p}_{\mathrm{head}}=1$ if $N=1$ is clearly not
of the same standing as saying $\hat{p}_{\mathrm{head}}=1$ after $N=10,000$
tosses. 
This invites us to look for a point estimator that itself reflects our
changing level of confidence as $N$ changes. 

Another peculiarity of the ML estimator is visible in Fig.~\ref{fig:llhN}: 
The point estimator is reported as a point on the boundary of the allowed
values for $p_{\mathrm{head}}$. 
This corresponds to the statement that tails can never occur. 
In general, rank deficiency in the estimator---that is, there exists at least
one pure state $|\psi\rangle$ on which $\hat{\rho}_{\mathrm{ML}}$ has no
support, $\langle \psi|\hat{\rho}_{\mathrm{ML}}|\psi\rangle=0$---says that a
detector that projects into the rank-deficient sector can \emph{never} click,
a statement that cannot be justified with finite $N$. 
Yet, the ML estimator is frequently rank-deficient whenever $N$ is of the
order of $K$, such that there is non-negligible probability that at least one
of the detectors has no clicks. 
This invites us to look for a point estimator that is full-rank for all finite
$N$.

\subsection{Mean estimators}\label{sec:meanEst}
An alternative to the ML estimator that takes $N$ into account is suggested by
Fig.~\ref{fig:llhN}: 
For small $N$, the likelihood is significant for a large region of
$p_{\mathrm{head}}$ values around the maximum; for large $N$, the likelihood
rapidly drops as we move away from the maximum. 
This suggests using the likelihood function as a \emph{weight} to construct a
point estimator. 
We weigh each state $\rho$ of the system by its likelihood, given data $D_N$,
and perform an average over all states to obtain the \emph{mean estimator} 
\begin{equation}\label{eq:meanEst}
\rhoEst(D_N)\equiv 
\frac{\ds\int \mathrm{d}\phi(\rho)\,\cL{\left(D_N|\rho\right)}\,\rho}
{\ds\int \mathrm{d}\phi(\rho)\,\cL{\left(D_N|\rho\right)}},
\end{equation}
where $\mathrm{d}\phi$ is an integration measure that tells us how to perform
a sum over states; 
$\mathrm{d}\phi$ should be non-negative on all physical states of the system,
and zero elsewhere. 
We can require, in addition, that $\int \mathrm{d}\phi(\rho)=1$ for
interpretation of $\mathrm{d}\phi$ as a probability distribution. 
This is, however, not necessary and we only require that $\mathrm{d}\phi$ is
not too pathological, so that the integrals in \eqref{eq:meanEst} exist. 

A reader familiar with Bayesian methods will recognize that the use of a
\emph{prior distribution} $\mathrm{d}\mu(\rho)$, which encapsulates the
experimenter's prior information about the probability of occurrence of each
state $\rho$, fits within this framework of mean estimators. 
If one chooses $\mathrm{d}\phi=\mathrm{d}\mu$, then by Bayes's theorem,
$\mathrm{d}\phi(\rho)\cL(D_N|\rho)$ is proportional to the 
\emph{posterior distribution} $\mathrm{d}\mu(\rho|D_N)$, that is, the
probability for state $\rho$ given data $D_N$. 
The mean estimator for this choice of $\mathrm{d}\phi$ is then simply the mean
state for the posterior distribution. 
This particular mean estimator has a long history in Bayesian estimation, and
is also sometimes used in the quantum literature. 

For us, the integration measure $\mathrm{d}\phi$ need not be chosen to
represent our prior information about the identity of the true state, but is a
functional parameter that we can adjust to satisfy desired optimality
conditions. 
By varying $\mathrm{d}\phi$, we can describe a reasonable class of point
estimators---the class of mean estimators---constructed as in
\eqref{eq:meanEst}, that is, 
\begin{equation}
\hat{\cS}_{\mathrm{ME}}\equiv \{\hat{\rho}_{\mathrm{ME}}[\mathrm{d}\phi]\}.
\end{equation}

For the case of a $K$-sided die, the mean estimator can be written more
explicitly as 
\begin{align}\label{eq:MEdie}
\hat{\rho}_{\mathrm{ME}}
&\equiv \sum_{k=1}^K\left(\hat{p}_k\right)_{\mathrm{ME}}\Lambda_k,\nonumber\\
\textrm{with}\ \pkEst
&\equiv \frac{\ds\int_0^1 (\mathrm{d}p)\,\delta{\left(1-\sum_{l=1}^Kp_l\right)} 
f(p)\cL{\left(D_N|p\right)}\,p_k}
{\ds\int_0^1 (\mathrm{d}p)\,\delta{\left(1-\sum_{l=1}^Kp_l\right)} 
f(p)\cL{\left(D_N|p\right)}}.
\end{align}
Here, we parameterize the states of the die by their 
probabilities $\{p_k\}$. 
The integration measure is written explicitly as $\mathrm{d}\phi(\rho)=%
(\mathrm{d}p)\,\delta{\left(1-\sum_{l=1}^Kp_l\right)}f(p)$, where
$(\mathrm{d}p)$ denotes the volume element 
$\mathrm{d}p_1\mathrm{d}p_2\ldots\mathrm{d}p_K$, the delta function enforces
$\sum_l p_l=1$, and $f(p)$ is a non-negative function that we can choose to
suit our needs.  

A natural symmetry in the $K$-sided die problem lies in the labeling of the
different faces as $1,2,\ldots,K$: 
A permutation of these arbitrary labels does not change the physical
description of the die.   
This symmetry should be reflected in the choice of $f(p)$ as an invariance
under permutation of the label $k$. 
A particularly simple choice of $f$ with this invariance is
\begin{equation}\label{eq:priorf}
f(p)= \left(\prod_{k=1}^Kp_k\right)^{\beta-1}\quad\textrm{with}\ \beta>0.
\end{equation}
The resulting mean estimator for this choice of $f$ is
\begin{equation}\label{eq:addbeta}
\left(\hat{p}_k\right)_{\mathrm{ME}}
=\frac{n_k+\beta}{N+K\beta}.
\end{equation}

There is an alternate way of arriving at this estimator following ML ideas. 
Suppose we obtained data $D_N\sim\{n_1,n_2,\ldots n_k,\ldots,n_K\}$. 
We add ``fake counts'' to every detector, of an amount $\beta$, so that the
data becomes 
$D_{N,\beta}\sim\{n_1+\beta,n_2+\beta,\ldots,n_k+\beta,\ldots,n_K+\beta\}$ 
and the total number of counts appears to be $N+K\beta$. 
Then, the ML estimator for this modified data is exactly that given in
\eqref{eq:addbeta}. 
This estimator is sometimes referred to as the ``add-$\beta$'' estimator, and
is used as an ad-hoc procedure to avoid reporting an ML estimator that lies on
the boundary. 
Note that the ${\beta\to0}$ limit of \eqref{eq:addbeta} corresponds exactly to
the ML estimator of \eqref{eq:MLdie}, but one cannot use the ${\beta=0}$
version of \eqref{eq:priorf} in \eqref{eq:MEdie}.

\subsection{Assessing the quality of an estimation procedure}
How well does a particular estimation procedure perform? 
To answer this question, we need to define a figure-of-merit that quantifies
how far from the true state an estimator is. 
Given data $D_N$, we compute the error in our guess $\hat{\rho}(D_N)$ of the
true state,  
\begin{equation}
\textrm{estimation error}
\equiv E(\hat{\rho},\rho,D_N)\equiv\mathrm{dist}(\hat\rho(D_N),\rho),
\end{equation}
where $\mathrm{dist}(\hat{\rho},\rho)$ is often chosen to be a formal distance
between two states $\rho$ and $\hat{\rho}$, like the trace distance or the
Euclidean distance. 
More generally, it is a function that assigns a (non-negative) ``cost''
whenever $\rho\neq \hat{\rho}$. 

Of course, we are not always going to get the data $D_N$ every time we perform
tomography. 
Instead, one should assess the efficacy of the estimation procedure for $\rho$
by averaging the estimation error over all possible data $D_N$ that one could
have obtained. 
Using terminology standard in estimation theory
(see, for example, Ref.~\refcite{LehmannCasella}), this gives the \emph{risk}
\begin{equation}\label{eq:Risk}
R_N(\hat{\rho},\rho)\equiv \sum_{D_N\in\cD_N}\cL(D_N|\rho)E(\hat\rho,\rho,D_N).
\end{equation}

The risk $R_N(\hat{\rho},\rho)$ still only tells us how good the estimator
$\hat{\rho}$ is for a given true state $\rho$. 
But, we have to judge the merits of an estimation procedure while not knowing
the identity of the true state (hence the need for tomography). 
If there exists a $\hat{\rho}$ such that the risk $R_N(\hat{\rho},\rho)$ for
\emph{all} true states $\rho$ is smaller than that of any other estimation
procedure, this $\hat{\rho}$ will clearly be the best procedure to use. 
However, an estimator with such miraculous properties is not likely to exist.

Instead, suppose we only ask that the estimator performs well ``on average''
over the true states. 
For example, a $\hat{\rho}$ that gives a large risk for a particular state
$\rho_0$ but small risk values for all other true states can be considered a
good estimation procedure as long as the probability that $\rho_0$ is indeed
the true state is tiny compared to other states. 
This requires some knowledge about the probability distribution of the true
states, that is, the prior distribution $\mathrm{d}\mu(\rho)$. 
If we know the prior distribution, a figure-of-merit that can be used to
assess an estimation procedure is its average performance over the true
states, that is, the risk weighted by the prior distribution, 
\begin{equation}
F_N(\hat{\rho},\mathrm{d}\mu)\equiv 
\int \mathrm{d}\mu(\rho)R_N(\hat{\rho},\rho).
\end{equation}
We refer to $F_N$ as the average risk. 
This includes the case where one does know the identity of the true state to
be some state $\tau$: $\mathrm{d}\mu(\rho)=\mathrm{d}\rho~\delta(\rho-\tau)$, 
so that $F_N(\hat{\rho},\mathrm{d}\mu)=R_N(\hat\rho,\tau)$. 
The case where $\hat{\rho}$ performs poorly only on a single state $\rho_0$
out of a possible (discrete) set of states $\cS\equiv \{\rho_i\}_{i=0}^L$
involves using the prior $\mu(\rho_i)=q_i$ for $\rho_i\in\cS$ and $0$
otherwise, with $q_0\ll q_{i\neq 0}$. 
A large risk for $\rho_0$ is suppressed in $F$ by a small enough value of $q_0$.

Given a prior distribution, the average risk quantifies the efficacy of the
estimator $\hat{\rho}$. To find the best estimation procedure among the set
$\hat\cS$ of all possible $\hat{\rho}$s, one minimizes the average risk: 
\begin{equation}
\min_{\hat{\rho}\in\hat\cS} F_N(\hat{\rho},\mathrm{d}\mu) \qquad\textrm{(Bayes)}.
\end{equation}
An estimator (not necessarily unique) that minimizes the average risk is known
as a \emph{Bayes estimator} for the prior distribution $\mathrm{d}\mu$. 
Note that this requirement of choosing a prior to assess the efficacy of an
estimation procedure in terms of average risk applies even for schemes like ML
methods which, by themselves, do not require a choice of prior or integration
measure.  

Bayes estimators are well studied in the state estimation literature. 
The following fact%
\footnote{This is Corollary 1.2 in Chapter 4 of Ref.~\refcite{LehmannCasella}.} 
relates mean estimators to Bayes estimators, which we will
find useful later (for a self-contained proof of this
fact, see \ref{app:Bayes0}): 
\begin{fact}\label{fact:Bayes0}
Suppose we choose the square of the Euclidean distance---the 
\emph{squared error}---to define the estimation error:
\begin{equation}
E(\hat\rho,\rho,D_N) =\mathrm{dist}\bigl(\hat{\rho}(D_N),\rho\bigr)
\equiv\tr{\left\{\bigl(\hat{\rho}(D_N)-\rho\bigr)^2\right\}}.
\end{equation}
Then, the unique Bayes estimator for prior distribution $\mathrm{d}\mu$ is the
mean estimator $\hat{\rho}_{\mathrm{ME}}[\mathrm{d}\mu]$. 
\end{fact}

The prior distribution $\mathrm{d}\mu(\rho)$ encapsulates our knowledge, not
of the source at hand, but of the preparer of the source. 
Imagine that the preparer, say Alice, has promised to provide a source that
puts out identical copies of a state $\rho$. 
Alice is, however, free to choose which particular state $\rho$ is. 
Our information about the probability that Alice provides us with a source
that puts out state $\rho$ is given by $\mathrm{d}\mu(\rho)$. 
The data $D_N$ collected from measuring a single instance of the source
provided by Alice cannot yield us any information about $\mathrm{d}\mu(\rho)$
(other than excluding states of the source that could not have given rise to
$D_N$). 
$\mathrm{d}\mu(\rho)$ must reflect prior knowledge about the preparer
gathered from previous interaction with different sources provided by Alice.  

In most tomographic scenarios, such prior knowledge is absent, and it seems
desirable to say ``we don't know''. 
Converting the heuristic notion of ``we don't know'' into a rigorous
``uninformative prior'' is, unfortunately, fraught with difficulties. 
For example, it would seem natural to assign equal probabilities to all
states, in the absence of knowledge of which states are more probable. 
This is not a problem for a discrete set of states labeled by a discrete label
$i$---it simply says that all the $q_i$s are equal. 
For a continuous set of states, which we parameterize using some continuous
parameter $x$, equal probability of occurrence means setting
$\mathrm{d}\mu(\rho)=dx$. 
However, there is no unique way of parameterizing the states, and equal
probability in one parameterization in general does not translate into equal
probability in a different parameterization. 
One often-used way to deal with this is to give up the idea of equal
probabilities for all states, and ask for an uninformative prior with the
property of parameterization invariance, for a relevant class of
parameterization. 
An example is the Jeffreys prior (see, for example, Ref.~\refcite{Jaynes},
p.~181 for a discussion), which is scale-invariant, that is, invariant under
reparameterization $x\rightarrow x^m$ for some power $m$.

\subsection{Minimaxity}
Assessing an estimator according to its average risk and using a Bayes
estimator for $\mathrm{d}\mu$ only works well if the true distribution
describing Alice the preparer is indeed $\mathrm{d}\mu$. 
Given that we do not usually know the prior distribution, and since even
choosing something like an uninformative prior is far from straightforward,
using a Bayes estimator for some choice of $\mathrm{d}\mu$ seems poorly
justified. 
Minimax approaches offer a way out of this.

Instead of using the average risk as a figure-of-merit, an alternative is to
use the worst-case risk, that is, the maximum risk (over all possible true
states) of using estimator $\hat{\rho}$. 
This does away with the requirement of choosing a prior distribution to
perform the averaging of the risk. 
The best estimator is found by minimizing the worst-case risk:
\begin{equation}
\min_{\hat{\rho}\in\hat{\cS}} \max_\rho R_N(\hat{\rho},\rho)\qquad\textrm{(minimax)}.
\end{equation}
An estimator (not necessarily unique) that minimizes the worst-case risk is
known as a \emph{minimax estimator}. 

Carrying out this double optimization to find a minimax estimator is, of
course, non-trivial. 
There is, however, a fact% 
\footnote{This is Corollary 1.5 in Chapter 5 of Ref.~\refcite{LehmannCasella}.} 
that can sometimes simplify the search for a minimax
estimator (see \ref{app:Bayes} for a self-contained proof):
\begin{fact}\label{fact:Bayes}
An estimator with constant risk that is also a Bayes estimator for some prior
distribution $\mathrm{d}\mu$ is a minimax estimator. 
If the estimator is also the unique Bayes estimator for some $\mathrm{d}\mu$,
then it is the unique minimax estimator. 
\end{fact}
\noindent Relating minimaxity to Bayes estimators is useful because much more
is known about Bayes estimators than minimax estimators. 
For example, Facts \ref{fact:Bayes0} and \ref{fact:Bayes} tell us that in
scenarios where the squared error is the suitable figure-of-merit, the unique
minimax estimator can be found by looking for a mean estimator
$\hat{\rho}\in\hat{\cS}_{\mathrm{ME}}$ with constant risk (if it exists). 
We will make use of this in the next section.

\subsection{The minimax estimator for the $K$-sided die}
For tomography with SIC-POMs, the squared error has the simple form
\begin{equation}\label{eq:sqErr}
\tr{\left\{\bigl(\hat{\rho}(D_N)-\rho\bigr)^2\right\}}
=\frac{(K-1)K}{(d-1)d}\sum_k(\hat{p}_k-p_k)^2,
\end{equation}
where $\hat{\rho}\equiv \sum_k \hat{p}_k\Lambda_k$ and 
$\rho\equiv \sum_k p_k\Lambda_k$. 
The squared error is proportional to the sum of squares of the difference in
the probabilities $\{\hat{p}_k\}$ and $\{p_k\}$. 
The corresponding risk is thus nothing more than the \emph{mean squared error}
(MSE) commonly used in classical state estimation. 

It is easy to work out the expression for the MSE for the class of mean
estimators for the $K$-sided die given in \eqref{eq:addbeta}. 
In particular, there exists a special value of $\beta$ (as used in
\eqref{eq:addbeta}) such that the MSE is independent of $p$, that is,
\emph{constant} over all states, 
\begin{equation}
\beta=\frac{\sqrt{N}}{K}.
\end{equation}
Using Facts \ref{fact:Bayes0} and \ref{fact:Bayes}, we know that the mean
estimator given in \eqref{eq:addbeta} with this value of $\beta$ is also
the unique minimax estimator for the $K$-sided die problem, with the choice of
the squared error as the figure-of-merit. 
This minimax property justifies objectively the choice of the integration
measure $\mathrm{d}\phi\equiv (\mathrm{d}p)\,f(p)=
(\mathrm{d}p)\,\Bigl(\prod_k p_k\Bigr)^{\beta-1}$ with $\beta=\sqrt{N}/K$. 

For this choice of $\beta$, we can write the minimax estimator
$\hat{\rho}_{\mathrm{MM}}\equiv
\sum_k\left(\hat{p}_k\right)_{\mathrm{MM}}\Pi_k$ for the $K$-sided die in a
form that exhibits its structure clearly, 
\begin{align}
\left(\hat{p}_k\right)_{\mathrm{MM}}&=\frac{1}{K}a_N+\nu_kb_N,\label{eq:pMM}
\nonumber\\
\textrm{with}\quad a_N&\equiv \frac{1}{1+\sqrt{N}},\quad 
b_N\equiv \frac{1}{1+1/\sqrt{N}}.
\end{align}
The parameters $a_N$ and $b_N$ depend only on $N$ and satisfy the relation
$a_N+b_N=1$. 
Observe that $a_N$ approaches zero as $N$ gets large, while $b_N$ approaches
unity, for which the minimax estimator approaches the ML estimator
$\left(\hat{p}_k\right)_{\mathrm{ML}}=\nu_k$. 
For $N$ small, $a_N$ is significant, and the two estimators differ.

Observe that, unlike the ML estimator, this minimax estimator is always
full-rank for finite $N$, since for any pure state $|\psi\rangle$ of the
system,  
\begin{equation}
\langle\psi|\hat\rho_{\mathrm{MM}}|\psi\rangle
=\frac{1}{K}a_N+b_N\sum_k\nu_k\vert\langle\psi|k\rangle\vert^2>0.
\end{equation}
We can also compute the purity of the minimax estimator,
\begin{align}
\tr{\left\{\hat\rho_{\mathrm{MM}}^2\right\}}
&=\sum_{k=1}^K \left(\hat p_k\right)_{\mathrm{MM}}^2
=\frac{1}{K}+b_N^2{\left(\sum_k\nu_k^2-\frac{1}{K}\right)}\nonumber\\
&\leq 1-{\left(1-\frac{1}{K}\right)}(1-b_N^2), \label{eq:diePurity}
\end{align}
which is strictly less than $1$ for finite $N$. 
The equality is attained when the data is such that all clicks are in a single
detector. 
Having the purity bounded away from $1$ is immediately obvious from the fact
that the estimator is always full-rank. 
However, the expression for the purity reveals more interesting features. 
For $K$ fixed, as $N$ increases, the bound on the purity increases towards
$1$, and the estimator can approach a pure state, expressing our increasing
confidence in claiming a definite pure state as we gather more data. 
Also, for $N$ fixed, the purity of the estimator decreases as $K$ increases. 
This is also intuitive: If $K$ is large, we would require more data to
convince ourselves that certain detectors will never click. 

The minimax estimator for the $K$-sided die problem circumvents both
complaints we had about the ML estimator. 
The minimax estimator itself has a dependence on $N$, and furthermore is never
rank-deficient for finite $N$. 
In the remainder of this paper, we would like to adapt this minimax estimator
to the quantum problem, while still retaining these two desirable properties.

\section{The quantum problem}\label{sec:Quantum}
In this section, we turn to the tomography of a quantum system. 
We begin by pointing out the differences between the classical and the quantum
problems (Sections \ref{subsec:SICPOM} and \ref{subsec:PhysConstr}). 
These considerations provide clues to adapting the minimax estimator of the
classical die problem to the quantum context (Sections \ref{subsec:PhysEst}
and \ref{subsec:FullRank}).

\subsection{SIC-POM for a quantum system}\label{subsec:SICPOM}
In moving from the classical to the quantum problem, the first difference we
meet is that the IC-POM that one can perform on the quantum system for full
tomography is non-unique. 
This has to do with the fact that there is no unique preferred basis such that
all quantum states are diagonal in that basis. 
We can, however, still choose to make use of a SIC-POM which offers efficiency
advantages over other choices of IC-POM~\cite{Renes04}. 
Related to the lack of a unique preferred basis is the fact that, unlike the
classical case, the POM outcomes of a SIC-POM are no longer mutually
orthogonal: 
$\tr{\left\{\Pi_k\Pi_l\right\}}\neq0$ for $k\neq l$. 
According to \ref{app:Geometry}, a SIC-POM for a quantum system has ${K=d^2}$,
and the $\Lambda_k$ operators of (\ref{eq:Lambda}) take the simple
form of $\Lambda_k=d(d+1)\Pi_k-1$. 
In our discussion below, we will only consider such a SIC-POM for tomography
of a quantum system. 

For a single qubit, that is, a two-dimensional quantum system, the SIC-POM is
the \emph{tetrahedron measurement}~\cite{Rehacek04}, with POM outcomes
proportional to projectors onto the legs of a regular tetrahedron inscribed
within the Bloch sphere. 
The tetrahedron measurement is non-unique in that the orientation of the
tetrahedron within the Bloch sphere is left to the choice and convenience of
the experimenter. 
Nevertheless, given a particular orientation, the POM outcomes of the
tetrahedron measurement can be written in terms of the Pauli vector operator
$\vec{\sigma}\equiv (\sigma_x,\sigma_y,\sigma_z)$ as 
\begin{equation}\label{eq:tetraPOM}
\Pi_k=\frac{1}{4}\bigl(1+\vec{a}_k\cdot \vec{\sigma}\bigr),\quad k=1,2,3,4,
\end{equation}
where each $\vec{a}_k$ is one of the four legs of the tetrahedron. 
The tetrahedron vectors $\vec{a}_k$ satisfy 
$\vec{a}_k\cdot\vec{a}_l=\frac{4}{3}\delta_{kl}-\frac{1}{3}$. 
Their linear dependence is captured by the facts that they sum to zero, 
$\sum_k \vec{a}_k=0$, and are complete, 
$\frac{3}{4}\sum_k\vec{a}_k\vec{a}_k=\dyadic{1}$. 
The probability of obtaining the $k$th outcome $\Pi_k$ for a qubit state 
$\rho=\frac{1}{2}\bigl(1+\vec{s}\cdot\vec{\sigma}\bigr)$ is given by 
$p_k\equiv \tr\{\rho\Pi_k\}=\frac{1}{4}\bigl(1+\vec{a}_k\cdot\vec{s}\bigr)$,
and $\Lambda_k=6\Pi_k-1$. 

Before we describe the quantum problem further, let us make a side remark
regarding the choice of figure-of-merit. 
For the classical die problem, while the state of the die can be gathered into
a single operator $\rho=\sum_k p_k\Lambda_k$, $\rho$ is but a book-keeping
device for the probabilities $\{p_k\}$ one is truly concerned with. 
The squared error, which directly measures how much the estimated
probabilities differ from the true probabilities, is thus a natural way to
quantify the estimation error. 
Of course, one can use the Euclidean distance (rather than its square), but
taking the square has analytical advantages. 
One just needs to note that doubling the difference in probabilities
quadruples the squared error. 

For the quantum problem, however, things are different. 
If the purpose of the quantum tomography is to predict the outcome of a future
measurement of the same SIC-POM used to perform tomography, then the $p_k$s
are again the only quantities of relevance, and the use of the squared error
is, as in the classical case, rather natural. 
However, if one's goal for tomography is to predict outcomes of a different
measurement that can yield information complementary to that provided by the
tomographic SIC-POM, then quantifying the estimation error in terms of
differences in the probabilities $\{p_k\}$ may not be suitable. 
Instead, one might choose to use, for example, the fidelity or the trace
distance between the estimator and the true state.%
\footnote{Note, however, that for the qubit problem, the square of the trace
  distance is equal to the squared error.} 
Nevertheless, for calculational ease, in the remainder of the paper, we shall
continue to use the squared error as our figure-of-merit. 

The squared error is also the (square of the) Euclidean distance between the
estimator and true state when viewed as vectors in the Hilbert-Schmidt space. 
We emphasize that there is no single figure-of-merit that is suitable for
all situations, but it should be chosen in accordance with the task at hand.

\subsection{Physicality constraints}\label{subsec:PhysConstr}
Consider any probability distribution. 
Does $\{p_k\}$ always correspond to outcome probabilities that can be obtained
by applying Born's rule for a SIC-POM to a physical state of the system? 
Equivalently, one can ask whether $\rho\equiv \sum_k p_k\Lambda_k$, for the
SIC-POM we are considering, describes a physical state of the system, that is,
$\rho$ has unit trace and is non-negative, for any probability distribution
$\{p_k\}$.

To answer this question, let us examine the quantity 
${p^2\equiv \sum_kp_k^2}$. 
Since ${\sum_kp_k=1}$, the minimum value of $p^2$ is attained when all the
$p_k$s are equal. 
This gives
\begin{equation}
p^2\geq\frac{1}{K},
\end{equation}
true for any probability distribution $\{p_k\}$.

What about the maximum value of $p^2$? 
Any physical state, whether quantum or classical, must satisfy
$\tr{\left\{\rho^2\right\}}\leq 1$. 
Using $\rho=\sum_kp_k\Lambda_k$, and writing $\Lambda_k=a\Pi_k+b$ for a
SIC-POM ($a$ and $b$ can be deduced from \eqref{eq:Lambda}), 
it is easy to show that $\tr{\left\{\rho^2\right\}}\leq 1$ implies 
$p^2\leq (1-b)/a$. 
A $K$-sided classical die problem has $a=1$ and $b=0$, which gives
\begin{equation}
\frac{1}{K}\leq \left(p^2\right)_{\mathrm{K-sided\ die}}\leq 1.
\end{equation}
This is satisfied for \emph{any} probability distribution $\{p_k\}$. 
The physicality requirement that $\tr{\left\{\rho^2\right\}}\leq 1$ does not
constrain the $p_k$s further. 
In fact, for the classical problem, $\rho=\sum_kp_k\Lambda_k$ is physical for
\emph{any} probability distribution.   

For the quantum problem with a SIC-POM, however, we have a different
situation. 
In this case, $a=d(d+1)$ and $b=-1$, which gives
\begin{equation}\label{eq:quantumConstraint}
\frac{1}{K}\leq\left(p^2\right)_{\mathrm{quantum}}\leq \frac{2}{d(d+1)}.
\end{equation}
The right side of the inequality is strictly less than $1$ for $d>1$. 
For example, the qubit problem with the tetrahedron measurement has the
physicality constraint 
\begin{equation}\label{eq:qubitConstraint}
\frac{1}{4}\leq \left(p^2\right)_{\mathrm{qubit}}\leq \frac{1}{3}.
\end{equation}
Only probability distributions $\{p_k\}$ that obey
\eqref{eq:quantumConstraint} can correspond to a physical quantum state. 
For example, $\{p_1=1,p_{2,3,4}=0\}$ does \emph{not} correspond to a physical
qubit state. 
This is a direct manifestation of the non-orthogonality of the POM
outcomes comprising a quantum SIC-POM. 
Note that \eqref{eq:qubitConstraint} is also sufficient for the qubit
problem: Any probability distribution $\{p_k\}$ satisfying
\eqref{eq:qubitConstraint} corresponds to a physical qubit state. 
For higher-dimensional quantum system, there are additional physicality
constraints, apart from \eqref{eq:quantumConstraint}, that $\{p_k\}$ must
satisfy.

Additional physicality constraints on $\{p_k\}$ mean that the expression for
the mean estimator for a quantum state is not just the expression for the
classical die problem given in \eqref{eq:MEdie}. 
For example, in the qubit problem, the mean estimator is now 
\begin{align}\label{eq:MEqb}
\hat\rho_{\mathrm{ME}}&\equiv 
\sum_{k=1}^K\left(\hat{p}_k\right)_{\mathrm{ME}}\Pi_k
\nonumber\\
\textrm{with}\quad\pkEst&\equiv 
\frac{\ds\int_0^1 (\mathrm{d}p)\,\delta{\left(1-\sum_{l=1}^4p_l\right)} 
\eta{\left(\frac{1}{3}-p^2\right)}f(p)\cL{\left(D_N|p\right)}\,p_k}
{\ds\int_0^1 (\mathrm{d}p)\,\delta{\left(1-\sum_{l=1}^4p_l\right)}
\eta{\left(\frac{1}{3}-p^2\right)} f(p)\cL{\left(D_N|p\right)}}.
\end{align}
Here, $\eta(\ )$ is Heaviside's unit step function: $\eta(x)=0$ if $x<0$,
and $\eta(x)=1$ if $x>0$. 
The step function enforces the upper bound in \eqref{eq:qubitConstraint}. 
With this expression, one can define an optimization procedure over all
possible $f(p)$ functions to look for a minimax (for example, using the MSE as
the risk) estimator for the qubit. 
This is, of course, difficult to perform. 
What would be simpler, is an $f(p)$ for which the MSE is constant for all
qubit states, which would then give a minimax estimator according to Facts
\ref{fact:Bayes0} and \ref{fact:Bayes}. 
Unfortunately, our preliminary attempts at this yielded a function $f(p)$ that
flies in the face of common sense.

\subsection{Adapting the classical minimax estimator to the quantum problem}
\label{subsec:PhysEst}
Instead of tackling the difficult problem of finding a minimax estimator, we
can try to build a simple estimator for quantum states by adapting the minimax
estimator from the classical die problem. 
For the quantum problem, the ML estimator has the same problem of having no
dependence on $N$ as well as suffering from rank deficiency. 
The ML estimator for the quantum case tells us to report, as the point
estimator, the physical quantum state at which the likelihood function attains
its maximum. 
For the qubit problem discussed above, this corresponds to looking for the
maximum of $\cL(D_N|\rho)$ subject not only to the usual constraint of $\sum_k
p_k=1$, but also the additional constraint that $\sum_k p_k^2\leq
\frac{1}{3}$. 
Whenever the data $D_N$ are such that $\sum_k\nu_k^2\leq \frac{1}{3}$, the ML
estimator is unmodified from the classical case: 
${\left(\hat{p}_k\right)}_{\mathrm{ML}}=\nu_k$; when the inequality is
violated, the ML estimator gives a state on the boundary of the Bloch sphere
(that is, a rank-deficient state) such that
$\sum_k{\left(\hat{p}_k\right)}_{\mathrm{ML}}^2=\frac{1}{3}$. 
The goal will be to adapt the minimax estimator from the classical die problem
in such a way that we arrive at an estimator that has a reasonable dependence
on $N$ and does not suffer from rank-deficiency. 

For the moment, let us put aside the desire for a full-rank estimator, and
focus on establishing a point estimator that is always physical for the
quantum problem. 
Suppose we begin with the expression for the $(\hat{p}_k)_{\mathrm{MM}}$s for
the $K$-sided die problem in \eqref{eq:pMM}, which we rename here as 
\begin{equation}\label{eq:unphysPk}
\hat{p}_{k,0}\equiv \frac{1}{K}a_N+\nu_kb_N.
\end{equation} 
We use these probabilities to construct an estimator for the quantum problem
with a $K$-outcome SIC-POM in accordance with \eqref{eq:invertBorn}, 
\begin{equation}\label{eq:unphysEst}
\hat{\rho}_0\equiv \sum_k \hat{p}_{k,0}\Lambda_k.
\end{equation}
We emphasize that the $\Lambda_k$s in \eqref{eq:unphysEst} are for the
quantum SIC-POM. 

Let us examine the quantity $p^2$ for $\{\hat{p}_{k,0}\}$, which we will
denote as $\hat{p}_0^2$. 
This was computed previously in \eqref{eq:diePurity}, which can be
rewritten as 
\begin{equation}
\hat{p}_0^2\leq\frac{2}{d(d+1)}
\left\{1+\frac{d-1}{2d}\left[\frac{(d+1)^2}{(1+1/\sqrt N)^2}-1\right]\right\}.
\end{equation}
Observe that this inequality is weaker, for $N\geq 1$ than the physicality
constraint $\hat{p}_0^2\leq \frac{2}{d(d+1)}$
in \eqref{eq:quantumConstraint}, necessary for $\hat{\rho}_0$ to be a
physical state. 
This means that there always exist data sets (for example, all but one
detector have zero clicks) for which $\hat{\rho}_0$ is not physical and,
therefore, fails to be a valid estimator for the quantum problem. 

Nevertheless, observe that data such that all but one detector have zero
clicks occur only with a small probability for values of $N$ that are not too
small. 
If the minimax estimator is physical for most data that we are likely to
encounter, then there is some hope that this estimator can still work well for
quantum states. 
After all, the quantum problem with a $K$-outcome SIC-POM looks very
similar---from the perspective of the outcome probabilities---to the classical
$K$-sided die problem as long as we are away from the boundary where the
physicality constraints come into play. 

Suppose we perform a ``correction'' to the estimator $\hat{\rho}_0$ whenever
it is unphysical, by admixing just enough of the maximally mixed state to make
the overall mixture physical. 
We take this new mixture to be the estimator
\begin{align}
\hat{\rho}&=(1-\lambda)\hat{\rho}_0+\frac{\lambda}{d}
\equiv \sum_k\hat{p}_k\Lambda_k,\nonumber\\
\textrm{with}\quad\hat{p}_k
&\equiv (1-\lambda)\hat{p}_{k,0}+\frac{\lambda}{d^2},\label{eq:QEst}
\end{align}
where $\lambda\geq 0$ is to be chosen as small as possible such that
$\hat{\rho}$ is a physical state. 
For the qubit problem, one can be more explicit: 
If $\hat{\rho}_0$ is a physical qubit state, $\lambda=0$; 
otherwise, $\lambda$ is chosen such that 
$\hat{p}^2\equiv \sum_k\hat{p}_k^2=\frac{1}{3}$. 
It follows that $\lambda$, for the qubit problem, is 
\begin{equation}
\lambda=\eta{\left( \sum_k\nu_k^2-\Aphy\right)}
{\left[1-\sqrt{\frac{\ds\Aphy-\frac{1}{4}}
                    {\ds\sum_k\nu_k^2-\frac{1}{4}}}\,\right]}.
\end{equation}
Here, $\Aphy$ is the largest value of $\sum_k\nu_k^2$ such that the data will
give a physical $\hat\rho_0$, that is, 
\begin{equation}
\sum_k \nu_k^2\leq\frac{1}{4}+\frac{1}{12 b_N^2}\equiv \Aphy,
\end{equation}
as implied by \eqref{eq:diePurity}.

Equation \eqref{eq:QEst} provides a simple prescription for converting any
estimator $\hat{\rho}_0$ from classical problems, not necessarily the minimax
estimator we have used here, into an estimator for the quantum problem. 
For example, one can get a good approximation to the ML estimator this way. 
Suppose we ignore physicality constraints and look for the ML estimator for
given data $D_N$ subject only to the constraints that the $p_k$s are
non-negative with unit sum. 
This is exactly the ML estimator if the problem is classical. 
We take this classical ML estimator for $\hat{\rho}_0$, and apply the
prescription given in \eqref{eq:QEst}. 
This gives an estimator that is very close (for example, in terms of fidelity)
to the one obtained from the ML scheme where one performs constrained
maximization (taking physicality constraints into account) of the likelihood
function.

\begin{figure}[t]
\includegraphics[width=\textwidth]{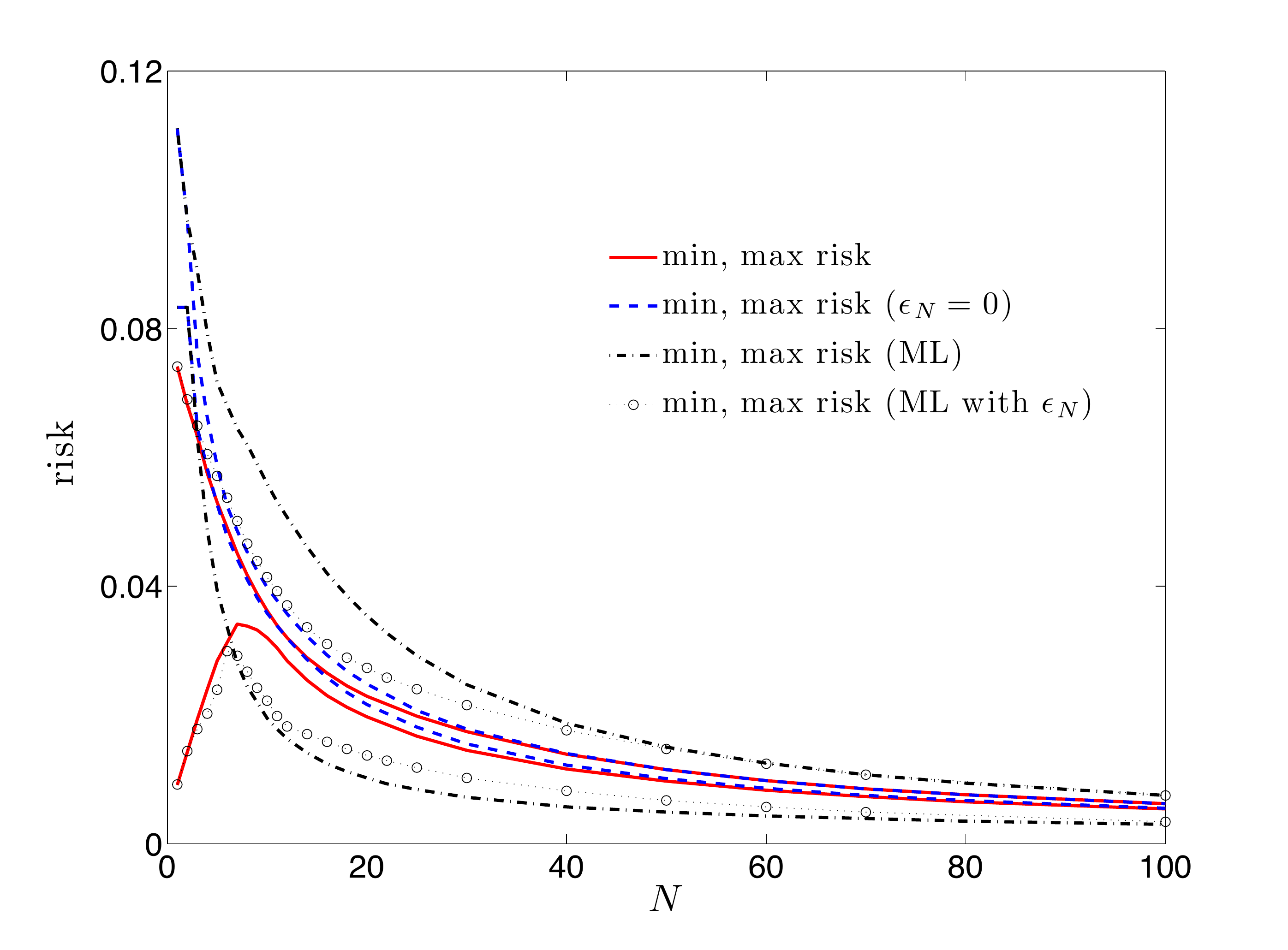}
\caption{\label{fig:RiskVsN}%
  The solid curves plot the minimum and maximum risk (over all qubit states) 
  for the optimal value of $\epsilon_N$. 
  The dash-dotted curves are the corresponding values for setting
  $\epsilon_N=0$ for all values of $N$. 
  The dotted curves correspond to the risk values for the ML estimator, and
  the curves with a circular marker give risk values for ML modified by an
  $\epsilon_N$ parameter as described in the text.} 
\end{figure}

To demonstrate the effectiveness of the estimator given in
\eqref{eq:QEst}, we plot in Fig.~\ref{fig:RiskVsN} the maximum and minimum
(over all possible input states) risks---measured using the MSE---for our
estimator in the qubit case 
(labeled in Fig.~\ref{fig:RiskVsN} as `min, max risk ($\epsilon_N=0$)'). 
For comparison, we also plot the corresponding risk values for the ML
estimator (labeled in Fig.~\ref{fig:RiskVsN} as `min, max risk (ML)'). 
Observe that the maximum error for our estimator is significantly smaller than
that for the ML estimator, indicating a step closer towards a true minimax
estimator. 
For $N$ not too small, note also the much smaller difference between the
maximum and minimum risk values for our qubit estimator as compared to the
case for the ML estimator.  
This near-homogeneity of risk values over all states is inherited from the
original classical minimax estimator that has constant risk. 
Risk homogeneity is attractive since it reflects an equal treatment of all 
input states, without having to implement the subjective construct of a
uniform prior for a continuous set of states.

\subsection{Modifying the estimator to be full-rank---a minimax estimator}
\label{subsec:FullRank}
In correcting the minimax estimator from the classical problem for physicality
in the quantum case, we have lost the feature that the resulting estimator is
always full-rank: 
Whenever $\hat{\rho}_0$ is unphysical, we correct it by choosing $\lambda$
just large enough to exactly cancel the most negative eigenvalue of
$\hat\rho_0$, and so to give a non-negative state with (at least) one zero
eigenvalue. 
In this section, we attempt to remedy this rank deficiency using a minimax
approach. 

Let us focus on the qubit problem with the tetrahedron measurement. 
We consider the same estimator as before in \eqref{eq:QEst} with $d=2$. 
Now, rather than choosing $\lambda$ such that the physicality constraint on
$\hat{p}^2$ is saturated ($=\frac{1}{3}$), we choose $\lambda$ such that we
saturate the constraint except for an overall factor of $(1-\epsilon_N)$, for
some parameter $\epsilon_N\geq 0$. 
More precisely, we set $\lambda=0$ whenever
$\hat{p}_0^2\leq\frac{1}{3}(1-\epsilon_N)$; 
otherwise, $\lambda$ is chosen to ensure $\hat p^2=
\frac{1}{3}(1-\epsilon_N)$. 
%This condition 
The latter case can be written more explicitly as
an equation for $\lambda$,
\begin{equation}\label{eq:lambda}
(1-\lambda)^2 b_N^2{\left(\sum_k\nu_k^2-\frac{1}{4}\right)}
=\frac{1-4\epsilon_N}{12}.
\end{equation}

What remains is to choose the value of $\epsilon_N$. For this, we make use of
a minimax procedure: 
Find the best value of $\epsilon_N$ by minimizing the worst-case risk, that
is, 
\begin{equation}\label{eq:minimaxEpsilon}
\min_{\epsilon_N\geq 0}\max_{\rho}R_N{\left(\hat{\rho},\rho\right)},
\end{equation}
where $\hat{\rho}$ is the estimator constructed from \eqref{eq:QEst} with
$\lambda$ chosen (when necessary) to satisfy \eqref{eq:lambda}. 
Equations \eqref{eq:QEst} and \eqref{eq:lambda} together define a class of
estimators $\hat{\cS}_{\epsilon_N}$ parameterized by $\epsilon_N$. 
The solution of the optimization problem stated in
\eqref{eq:minimaxEpsilon} is a minimax estimator in the restricted class
$\hat{\cS}_{\epsilon_N}$ of estimators. 

\begin{figure}[t]
\begin{center}
\includegraphics[width=0.8\textwidth]{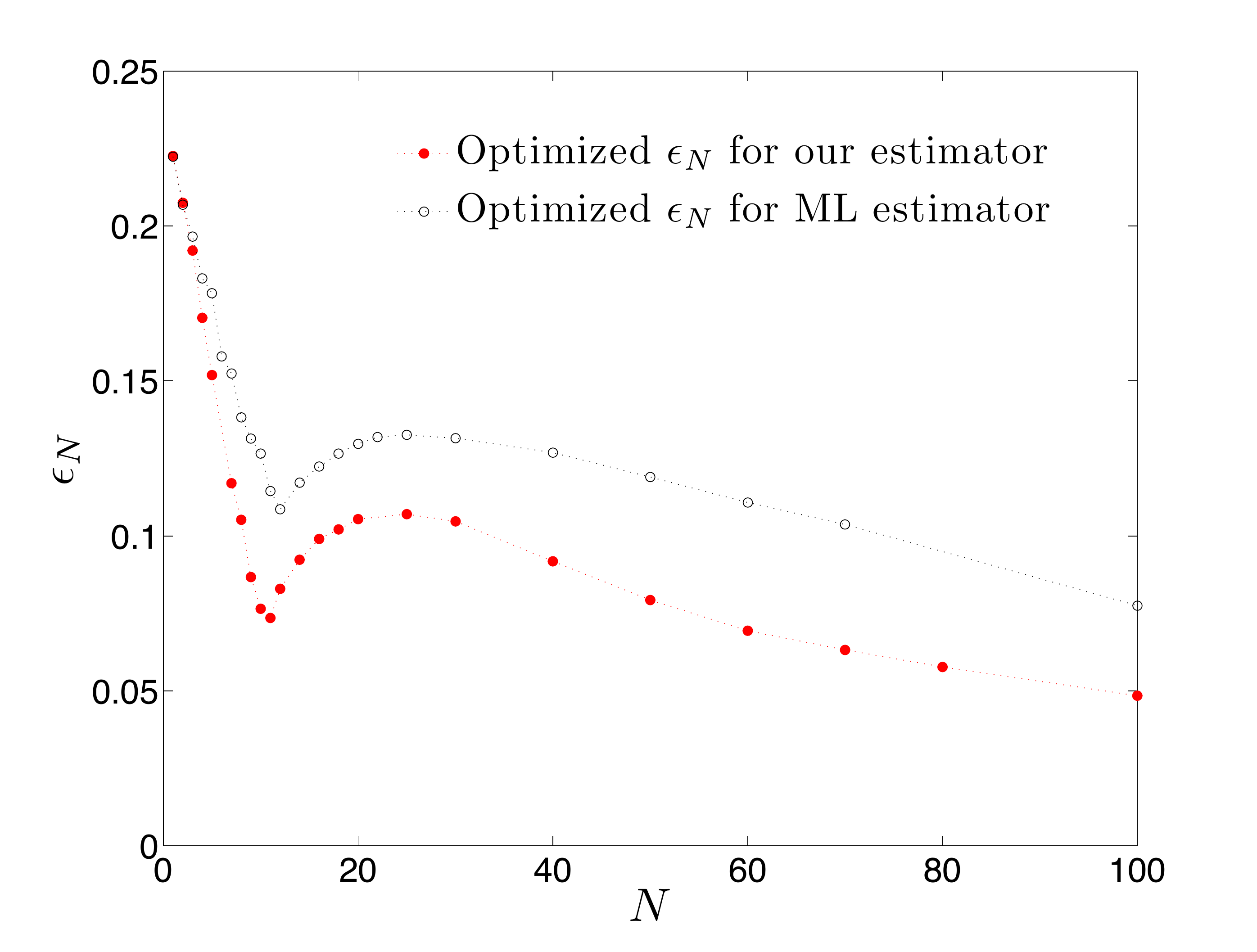}
\caption{\label{fig:EpVsN}%
  Optimized value of $\epsilon_N$ for the qubit problem for different values
  of $N$, with $\hat\rho_0$ defined as in \eqref{eq:unphysPk} and
  \eqref{eq:unphysEst}, as well as for the ML estimator modified by an
  $\epsilon_N$ parameter.} 
\end{center}
\end{figure}

Figure \ref{fig:EpVsN} reports the optimal values of $\epsilon_N$ as a
function of $N$, with $\hat{\rho}_0$ defined as in \eqref{eq:unphysPk}
and \eqref{eq:unphysEst} and restricted to the qubit case 
(labeled in Fig.~\ref{fig:EpVsN} 
as `Optimized $\epsilon_N$ for our estimator'). 
The performance of the estimator with the optimal value of $\epsilon_N$ is
plotted in Fig.~\ref{fig:RiskVsN} 
(labeled in Fig.~\ref{fig:RiskVsN} as `min, max risk'). 
Observe that as $N$ grows, the difference in performance between optimizing
the value of $\epsilon_N$ and choosing $\epsilon_N=0$ (that is, the estimator
discussed in Section \ref{subsec:PhysEst}) rapidly diminishes. 
If desired, for practical convenience, one can set $\epsilon_N=0$ for
$N\gtrsim 100$. 
That $\epsilon_N$ approaches $0$ as $N$ increases is particularly rewarding
because it is in line with our intuition that for small $N$, we have little
evidence that can support reporting a point estimator that is close to a
rank-deficient state; however, as $N$ increases, we gather more and more data
and gain confidence in reporting a state that is closer and closer to a
rank-deficient state, as described by our estimator with $\epsilon_N$
approaching zero.  

For comparison, we have also plotted the performance of the ML estimator, with
the modification that one restricts the domain of the maximization of the
likelihood function to states such that $p^2\leq\frac{1}{3}(1-\epsilon_N)$,
where $\epsilon_N$ is again chosen via the same minimax procedure as above. 
This simple modification removes the problem of rank-deficiency of the usual
ML estimator.
In fact, as can be seen from Fig.~\ref{fig:RiskVsN} 
(line labeled `min, max risk (ML with $\epsilon_N$)'), it
significantly improves the maximum risk for the ML estimator, although it does
not do nearly as well as the estimator discussed above. 

Our approach to a restricted minimax estimator can be extended beyond the
qubit case and beyond using a $\rho_0$ that comes from the classical minimax
estimator. 
As mentioned in Section \ref{subsec:PhysEst}, one can begin with one's
favorite classical estimator and admix enough of the completely mixed state to
ensure physicality of the resulting estimator. 
To fix the rank-deficiency problem, one can then use a similar minimax
procedure as in \eqref{eq:minimaxEpsilon} to find the best estimator that
avoids the physicality boundary. 
In the qubit case, a single parameter $\epsilon_N$ was sufficient to delineate
the physicality boundary and characterize the relevant class of estimators. 
For higher dimension, physicality constraints are more complicated (and, in
fact, are often not well understood), and one would typically require more
than one parameter to define the analog of $\hat{\cS}_{\epsilon_N}$. 
Nevertheless, the same minimax procedure is applicable.

Variants of our estimator are also possible.
For example, one can treat both $\epsilon_N$ as well as $b_N$ 
(with $a_N=1-b_N$) in \eqref{eq:unphysPk} as parameters that we choose in
a minimax fashion.  
Another variant, for the case of the tetrahedron POM for the qubit, 
suggests itself when we examine \eqref{eq:lambda} which 
determines $\lambda$ whenever $\hat p_0^2>\frac{1}{3}(1-\epsilon_N)$:
The choice ${b_N^2=1-4\epsilon_N}$ gives a particularly simple value for
$\lambda$ that depends only on the relative frequencies $\nu_k$, but not on
the parameters $b_N$ and $\epsilon_N$. 
One then performs minimax optimization over $\epsilon_N$ only. 
Both variants give results very similar to our estimator above for the case of
the tetrahedron POM for the qubit.

\section{Conclusion}
We demonstrated a simple procedure for adapting the minimax estimator for the
classical die problem to the quantum case of a single qubit with the
tetrahedron measurement. 
We obtained an estimator that inherited desirable properties from the
classical version: 
(i) It is always full rank and contains a reasonable $N$ dependence; 
(ii) it has much smaller maximum risk, as measured by the mean squared error,
compared to the popular ML estimator; 
(iii) it gives nearly constant risk over all states and hence treats all
possible states in a fair manner. 

The procedure of admixing a sufficient amount of the completely mixed state to
obtain a physical and full-rank estimator can be applied to any estimator
appropriate for the analogous classical problem. 
For typical data and most states, the classical estimator is usually physical;
it is only the rare case that requires a physicality correction. 
This automatically ensures that the resulting quantum estimator will inherit
most of the properties of the classical estimator. 
One can, for example, do this for estimators for the classical die problem
that are minimax for other risk functions (for example, based on relative
entropy).  
The procedure is also applicable beyond the qubit case and also beyond a
SIC-POM. 
For higher dimensions, the physicality constraints will involve more
inequalities that the probabilities $\{p_k\}$ must satisfy, but can, in
principle, be imposed as additional constraints for the choice of the admixing
parameter $\lambda$. 
In every case, a minimax procedure can be used to choose parameters like
$\epsilon_N$ to avoid the boundary. 
Note also that, for problems with an unusual symmetry, one can in fact
consider admixing not the completely mixed state but some other suitable
reference state.

Given the simplicity of this estimator, we believe it will find much utility
in tomographic experiments as a first-cut point estimate of the unknown
state. 
Future work exploring the effectiveness of this procedure for other
estimators, risk functions, and higher dimensions can also be potentially
interesting. 
Progress towards general minimax estimators following the programme set up in
this paper will also certainly be of importance to quantum tomography.

\section*{Acknowledgments}
We are grateful for insightful discussions with David Nott and Benjamin Phuah.
This work is supported by the National Research Foundation and the Ministry of
Education, Singapore.

\appendix

\section{Geometry of  S-POMs}%symmetric probability operator measurements}
\label{app:Geometry}
The $K$ outcomes $\Pi_k$ of an S-POM for a $d$-dimensional system obey 
\begin{eqnarray}
  \label{eq:C1}
  \tr\bigl\{\Pi_j\Pi_k\bigr\}&=&\frac{d}{K}\Bigl[w\delta_{jk}
                  +\frac{1-w}{K-1}(1-\delta_{jk})\Bigr],
\nonumber\\
  \tr\bigl\{\Pi_k\bigr\}&=&\frac{d}{K},
\end{eqnarray}
with
\begin{equation}
  \label{eq:C2}
  \frac{1}{K}\leq w\leq 1.
\end{equation}
The lower bound applies when the $\Pi_k$s are multiples of the identity, which
is a case of no interest; the upper bound applies when the outcomes have
support in pairwise orthogonal subspaces.
If the outcomes are (subnormalized) rank-$r$ projectors, we have
$w=d/(rK)$ with ${1\leq d/r\leq K}$.
Of particular importance is the rank-$1$ situation, for which
\begin{equation}
  \label{eq:C3}
  {\Pi_k}^2=\frac{d}{K}\Pi_k,\qquad
  \tr\bigl\{\Pi_j\Pi_k\bigr\}=\frac{d^2}{K^2}\Bigl[\delta_{jk}
                  +\frac{K-d}{(K-1)d}(1-\delta_{jk})\Bigr]
\end{equation}
hold.

The set of traceless hermitian operators constitute a real
$(d^2-1)$-dimensionless vector space that we endow with the 
Hilbert-Schmidt inner product
\begin{equation}
  \label{eq:C4}
  A\cdot B\equiv\tr\{AB\}\quad\mbox{for\ } A^\dagger=A,\ B^\dagger=B,\ 
\tr\{A\}=\tr\{B\}=0.
\end{equation}
Since the operators ${\Pi_k-1/K}$ are in this vector space, we can state
(\ref{eq:C1}) as
\begin{equation}
  \label{eq:C5}
  \Bigl(\Pi_j-\frac{1}{K}\Bigr)\cdot\Bigl(\Pi_k-\frac{1}{K}\Bigr)
=\frac{d}{K}\frac{wK-1}{K}\Bigl[\delta_{jk}-\frac{1}{K-1}(1-\delta_{jk})\Bigr].
\end{equation}
In conjunction with ${\sum_k(\Pi_k-1/K)=0}$, this tells us that the vectors
${\Pi_k-1/K}$ define a flat $K$-edged pyramid, if we employ the terminology of
Ref.~\refcite{pyramids}. 
In the rank-$1$ situation of (\ref{eq:C3}), the prefactor in
(\ref{eq:C5}) is ${(d-1)d/K^2}$. 

In view of this geometrical property of the S-POM, there can be at most $d^2$
outcomes. 
Indeed, the S-POM is IC for ${K=d^2}$, but not when ${K<d^2}$,
and there are no S-POMs with ${K>d^2}$.
In an alternative way of reasoning, we represent the vectors
$\Pi_k-\frac{1}{K}$ by the columns of the $K\times K$ matrix
\begin{equation}
  \label{eq:C5a}
\sqrt{\frac{(wK-1)d}{(K-1)K^3}}
\left[
  \begin{array}{cccc}
    K-1 & -1 & \cdots & -1\\
     -1 & K-1 & \cdots & -1\\
    \vdots&\vdots&\ddots&\vdots\\
    -1 & -1 & \cdots & K-1
  \end{array}
\right]
\end{equation}
and note that this matrix has rank ${K-1}$, which implies that the
$K$ vectors $\Pi_k-\frac{1}{K}$ span a ${(K-1)}$-dimensional subspace.

Regarding the statistical operator $\rho$, we note that ${\rho-1/d}$ is
hermitian and traceless, and so are the operators ${\Lambda_k-1/d}$ 
that appear in (\ref{eq:invertBorn}), 
\begin{eqnarray}
  \label{eq:C6}
  \rho=\sum_{k=1}^Kp_k\Lambda_k
      =\frac{1}{d}
       +\sum_{k=1}^K\Bigl(p_k-\frac{1}{K}\Bigr)\Bigl(\Lambda_k-\frac{1}{d}\Bigr),
\end{eqnarray}
where either ${p_k-1/K\to p_k}$ or ${\Lambda_k-1/d\to\Lambda_k}$ is a
permissible replacement, but not both.
The defining property of the $\Lambda_k$s, namely
$\tr\{\Pi_j\Lambda_k\}=\delta_{jk}$ or
\begin{equation}
  \label{eq:C7}
  \Bigl(\Pi_j-\frac{1}{K}\Bigr)\cdot\Bigl(\Lambda_k-\frac{1}{d}\Bigr)
=\frac{K-1}{K}\Bigl[\delta_{jk}-\frac{1}{K-1}(1-\delta_{jk})\Bigr],  
\end{equation}
implies their standard form,
\begin{eqnarray}
  \label{eq:C8}
  \Lambda_k&=&\frac{1}{d}
              +\frac{(K-1)K}{(wK-1)d}\Bigl(\Pi_k-\frac{1}{K}\Bigr)
\nonumber\\
  &=&\frac{(K-1)K}{(wK-1)d}\Pi_k-\frac{(1-w)K}{(wK-1)d}.
\end{eqnarray}
If the S-POM is not IC (${K<d^2}$), 
the $\Lambda_k$s are not uniquely determined, because there is then 
the option to add a traceless hermitian operator on the right-hand side 
of (\ref{eq:C8}) that is orthogonal to all $K$ vectors ${\Pi_k-1/K}$.
It follows that the statistical operator $\rho$ of (\ref{eq:C6}) is not 
unique unless the S-POM is a SIC-POM.
What \emph{is} unique, however, is the part of ${\rho-1/d}$ that resides in
the $(K-1)$-dimensional subspace spanned by the vectors ${\Pi_k-1/K}$. 

For the standard $\Lambda_k$s of (\ref{eq:C8}), 
the vectors ${\Lambda_k-1/d}$ make up the same flat pyramid as the vectors
${\Pi_k-1/K}$, except that the edges have different lengths.
More specifically, we have
\begin{equation}
  \label{eq:C9}
   \Bigl(\Lambda_j-\frac{1}{d}\Bigr)\cdot\Bigl(\Lambda_k-\frac{1}{d}\Bigr)
  =\frac{(K-1)^2}{(wK-1)d}
   \Bigl[\delta_{jk}-\frac{1}{K-1}(1-\delta_{jk})\Bigr],
\end{equation}
and
\begin{equation}
  \label{eq:C10}
  \frac{K}{\sqrt{(wK-1)d}}\Bigl(\Pi_k-\frac{1}{K}\Bigr)
=\frac{\sqrt{(wK-1)d}}{K-1}\Bigl(\Lambda_k-\frac{1}{d}\Bigr)
\end{equation}
are the edge vectors of the generic pyramid with unit-length edges.

In the qubit case (${d=2}$), the ${(d^2-1)}$-dimensional real vector space of
traceless hermitian operators is isomorphic to the three-dimensional cartesian
space in which the Bloch ball is embedded. 
Rank-$1$ outcomes are of the form
\begin{equation}
  \label{eq:C11}
  \Pi_k=\frac{1}{K}\Bigl(1+\vec{e}_k\cdot\vec{\sigma}\Bigr)
\end{equation}
with ${K=2}$ for the von Neumann measurement, ${K=3}$ for the so-called trine
measurement, and ${K=4}$ for the tetrahedron measurement of
(\ref{eq:tetraPOM}). 
The $\vec{e}_k$s are unit vectors, with 
\begin{equation}
  \label{eq:C12}
  \Bigl(\Pi_j-\frac{1}{K}\Bigr)\cdot\Bigl(\Pi_k-\frac{1}{K}\Bigr)
  =\frac{2}{K^2}\vec{e}_j\cdot\vec{e}_k
\end{equation}
stating how the inner product in the operator space is related to the scalar
product of three-dimensional vectors. 
When representing the $\vec{e}_k$s by three-component columns of cartesian
coordinates, possible choices are
\begin{eqnarray}
  \label{eq:C13}
  &\Bigl[\vec{e}_1\ \vec{e}_2\Bigr]
={\left[
    \begin{array}{cr}
      0&0 \\ 0&0 \\ 1&-1
    \end{array}
  \right]}\quad\mbox{for $K=2$,}\quad
  \Bigl[\vec{e}_1\ \vec{e}_2\ \vec{e}_3\Bigr]
=\ds\frac{1}{\sqrt{6}}{\left[
    \begin{array}{rrr}
      2&-1&-1 \\ -1&2&-1 \\ -1&-1&2
    \end{array}
  \right]}\quad\mbox{for $K=3$,}&\nonumber\\
&\mbox{and}\quad
  \Bigl[\vec{e}_1\ \vec{e}_2\ \vec{e}_3\ \vec{e}_4\Bigr]
=\ds\frac{1}{\sqrt{3}}{\left[
    \begin{array}{rrrr}
      1&-1&-1&1 \\ -1&1&-1&1 \\ -1&-1&1&\phantom{-}1
    \end{array}
  \right]}\quad\mbox{for $K=4$.}&
\end{eqnarray}
In each case, one easily confirms that $\sum_k\vec{e}_k=0$ and
\begin{equation}
  \label{eq:C14}
  \vec{e}_j\cdot\vec{e}_k
  =\delta_{jk}-\frac{1}{K-1}(1-\delta_{jk}),
\end{equation}
as implied by (\ref{eq:C12}) with (\ref{eq:C5}).

\section{Proof of Fact \ref{fact:Bayes0}}\label{app:Bayes0}
\noindent\textbf{Fact \ref{fact:Bayes0}.} 
Suppose we choose the square of the Euclidean distance to define the
estimation error: 
\begin{equation}
E(\hat{\rho},\rho,D_N) 
=\mathrm{dist}\bigl(\hat{\rho}(D_N),\rho\bigr)
\equiv\tr{\left\{\bigl[\hat{\rho}(D_N)-\rho\bigr]^2\right\}}.
\end{equation}
Then, the unique Bayes estimator for prior distribution $\mathrm{d}\mu$ is the
mean estimator $\hat{\rho}_{\mathrm{ME}}[\mathrm{d}\mu]$. 

\medskip

\noindent{\textbf{Proof.}} 
Consider any estimator $\hat{\rho}\in\hat{\cS}$. 
Inserting 
$0=-\hat{\rho}_{\mathrm{ME}}[\mathrm{d}\mu]+\hat{\rho}_{\mathrm{ME}}[\mathrm{d}\mu]$ 
into the squared Euclidean distance, the estimation error can be written as 
\begin{eqnarray}
E(\hat{\rho},\rho,D_N)&=&\tr\bigl\{[\rho-\hat{\rho}_{\mathrm{ME}}(D_N)]^2
+[\hat{\rho}_{\mathrm{ME}}(D_N)-\hat{\rho}(D_N)]^2\nonumber\\
&&\hphantom{\tr\bigl\{}
-2[\rho-\hat{\rho}_{\mathrm{ME}}(D_N)]
[\hat{\rho}_{\mathrm{ME}}(D_N)-\hat{\rho}(D_N)]\bigr\}.   
\end{eqnarray}
The average risk can then be computed as 
\begin{align}
F_N(\hat{\rho}, \mathrm{d}\mu)
&=\int\mathrm{d}\mu(\rho)\sum_{D_N\in\cD_N}\!\!
\cL(D_N|\rho)\,\tr\bigl\{[\rho-\hat{\rho}_{\mathrm{ME}}(D_N)]^2\bigr\}
\\&\quad
+\int\mathrm{d}\mu(\rho)\sum_{D_N\in\cD_N}\!\!
\cL(D_N|\rho)\,\tr\bigl\{[\hat{\rho}_{\mathrm{ME}}(D_N)-\hat{\rho}(D_N)]^2\bigr\}
\nonumber\\&\quad
-2\int \mathrm{d}\mu(\rho)\sum_{D_N\in\cD_N}\!\!
\cL(D_N|\rho)\,\tr\bigl\{[\rho-\hat{\rho}_{\mathrm{ME}}(D_N)]
[\hat{\rho}_{\mathrm{ME}}(D_N)-\hat{\rho}(D_N)]\bigr\}.
\nonumber
\end{align}
The third term is zero, by definition of 
$\hat{\rho}_{\mathrm{ME}}[\mathrm{d}\mu]$. 
The first term does not depend on $\hat{\rho}$ and just gives a fixed constant
value. 
To find the Bayes estimator, we thus solve the optimization problem
\begin{equation}
\min_{\hat{\rho}\in\hat{\cS}}\int \mathrm{d}\mu(\rho)\sum_{D_N\in\cD_N}\!\!
\cL(D_N|\rho)\tr{\left\{[\hat{\rho}_{\mathrm{ME}}(D_N)-\hat{\rho}(D_N)]^2\right\}},
\end{equation}
for which $\hat{\rho}=\hat{\rho}_{\mathrm{ME}}[\mathrm{d}\mu]$ is clearly the
unique solution.

\section{Proof of Fact \ref{fact:Bayes}}\label{app:Bayes}
\noindent\textbf{Fact \ref{fact:Bayes}.} 
An estimator with constant risk that is also a Bayes estimator for some prior
distribution $\mathrm{d}\mu$ is a minimax estimator. 
If the estimator is also the unique Bayes estimator for some $\mathrm{d}\mu$,
then it is the unique minimax estimator. 

\medskip

\noindent{\textbf{Proof.}} 
Suppose $\hat{\rho}_\mathrm{B}$ is a Bayes estimator for $\mathrm{d}\mu$ with
constant risk, that is, $R_N(\rho,\hat{\rho}_\mathrm{B})=R_{N,\mathrm{B}}$ for
all $\rho$ in $\cS$.  
$\hat{\rho}_\mathrm{B}$ satisfies
\begin{equation}
\int \mathrm{d}\mu(\rho)R_N(\rho,\hat{\rho}_\mathrm{B})
=R_{N,\mathrm{B}}=\max_{\rho\in\cS}R_N(\rho,\hat{\rho}_\mathrm{B}).
\end{equation}
Consider another estimator $\hat{\rho}\neq\hat{\rho}_\mathrm{B}$. 
Then, we have that
\begin{align}
\max_{\rho\in\cS}R_N(\rho,\hat{\rho})
&\geq \int \mathrm{d}\mu(\rho)R_N(\rho,\hat{\rho})\nonumber\\
&\geq \int \mathrm{d}\mu(\rho)R_N(\rho,\hat{\rho}_\mathrm{B})
=\max_{\rho\in\cS}R_N(\rho,\hat{\rho}_\mathrm{B}).\label{eq:app1}
\end{align}
The first inequality is simply a statement that the maximum is greater then
the mean; 
the second inequality follows from the fact that $\hat{\rho}_\mathrm{B}$ is a
Bayes estimator. 
Equation \eqref{eq:app1} says precisely that $\hat{\rho}_\mathrm{B}$ is minimax. 
If $\hat{\rho}_\mathrm{B}$ is also the \emph{unique} Bayes estimator for
$\mathrm{d}\mu$, the second inequality is converted into a strict inequality
(``$>$''), and we have $\max_{\rho\in\cS}R_N(\rho,\hat{\rho})
>\max_{\rho\in\cS}R_N(\rho,\hat{\rho}_\mathrm{B})$,
which proves the uniqueness of $\hat{\rho}_\mathrm{B}$ as a minimax estimator.

%\clearpage\input{loglabs}\input{bibcheck}

\end{document}

%%% Local Variables:
%%% mode: latex
%%% mode: latex-math
%%% mode: flyspell
%%% TeX-master: t
%%% End:
% LocalWords: